\documentclass[prl,twocolumn,showpacs,preprintnumbers,floatfix,superscriptaddress]{revtex4}

\usepackage{graphicx}
\usepackage{dcolumn}

\newcommand{\gev}{\,{\rm GeV}}
\newcommand{\non}{\nonumber}

\newcommand{\shat}{\hat{s}}
\newcommand{\eps}{\epsilon}

\newcommand{\GEp}{{G_E^p}}
\newcommand{\GEn}{{G_E^n}}
\newcommand{\GMp}{{G_M^p}}
\newcommand{\GMn}{{G_M^n}}
\newcommand{\GMs}{{G_M^s}}
\newcommand{\GEs}{{G_E^s}}
\newcommand{\GA}{G_A}

\newcommand{\GAEWp}{{\tilde{G}_A^p}}
\newcommand{\GAEWn}{{\tilde{G}_A^n}}
\newcommand{\GAEWN}{{\tilde{G}_A^N}}

\begin{document}

\preprint{
\hbox{JLAB-THY-06-479}}

\title{Extracting nucleon strange and anapole form factors from world data}

\author{R.~D.~Young}
\affiliation{Jefferson Lab,
             12000 Jefferson Ave.,
             Newport News, Virginia 23606, USA}
\author{J.~Roche}
\affiliation{Jefferson Lab,
             12000 Jefferson Ave.,
             Newport News, Virginia 23606, USA}
\affiliation{Rutgers,
             The State University of New Jersey,
             Piscataway, New Jersey 08854, USA}
\author{R.~D.~Carlini}
\author{A.~W.~Thomas}
\affiliation{Jefferson Lab,
             12000 Jefferson Ave.,
             Newport News, Virginia 23606, USA}

\begin{abstract}
The complete world set of parity violating electron scattering data up
to $Q^2\sim 0.3\gev^2$ is analysed. We extract the current
experimental determination of the strange electric and magnetic form
factors of the proton, as well as the weak axial form factors of the
proton and neutron, at $Q^2 = 0.1\gev^2$.  Within experimental
uncertainties, we find that the strange form factors are consistent
with zero, as are the anapole contributions to the axial form factors.
Nevertheless, the correlation between the strange and anapole
contributions suggest that there is only a small probability that
these form factors all vanish simultaneously.
\end{abstract}


\maketitle

%
%
Parity-violating electron scattering (PVES) is an essential tool in
mapping out the flavour composition of the electromagnetic form
factors. Exposing the role of the strange quark via these measurements
provides direct information on the underlying dynamics of
nonperturbative QCD --- a considerable achievement both experimentally
and theoretically.  The most precise separation of the strange
electric and magnetic form factors is available at $Q^2\simeq
0.1\gev^2$, where experiments by the SAMPLE
\cite{Ito:2003mr,Spayde:2003nr}, PVA4 \cite{Maas:2004dh} and HAPPEx
\cite{Aniol:2005zf,Aniol:2005zg} collaborations have been performed
with varying kinematics and targets. At higher $Q^2$, HAPPEx
\cite{Aniol:2000at,Aniol:2004hp}, PVA4 \cite{Maas:2004ta} and the
forward angle G0 experiment \cite{Armstrong:2005hs} provide further
information over the range $Q^2\sim 0.12$--$1.0\gev^2$.
Here we use systematic expansions of all the unknown form factors to
simultaneously analyze the current data set and extract the values at
$Q^2 = 0.1\gev^2$, independent of theoretical input --- other than the
constraint of charge symmetry. The results provide a critical test of
modern theoretical estimates of the anapole moment of the proton and
neutron as well as their strange form factors.

The proton-PVES experiments are sensitive to the strange form factors
$\GEs$ and $\GMs$, and the electroweak axial form factor $\GAEWp$ ---
which includes the anapole form factor
\cite{Musolf:1990ts,Zhu:2000gn}.  Previously, limited experimental
data made it difficult to carry out a simultaneous separation of all
three form factors; instead, assumptions were made on the
(in)significance of certain contributions based on the kinematic
domain and/or the use of theoretical calculations.  In combining
proton and deuteron data, there are two independent anapole form
factors. Together with the two strange form factors, this analysis
presents the first extraction of all four form factors from data. No
more than two independent terms have been fit simultaneously in any
previous analysis. Further, no analysis has attempted to determine the
isoscalar anapole term from data. This contribution is quite poorly
constrained by experiment and the design of an appropriate measurement
to improve this situation is both a theoretical and experimental
challenge.

%
%
The role of the strange quark is probed by measuring the PV asymmetry
in polarised $e$--$N$ scattering, for which the dominant contribution
arises from interference between the $\gamma$ and $Z^0$ exchange.  The
majority of measurements have been performed on hydrogen: SAMPLE
\cite{Spayde:2003nr,Beise:2004py}, HAPPEx
\cite{Aniol:2000at,Aniol:2005zg}, PVA4 \cite{Maas:2004ta,Maas:2004dh}
and G0 \cite{Armstrong:2005hs}.

As described in Ref.~\cite{Musolf:1993tb}, the PV
asymmetry for a proton target is given by (assuming charge symmetry)
\begin{eqnarray}
A_{LR}^p &=& \frac{d\sigma_R-d\sigma_L}{d\sigma_R+d\sigma_L}= 
-\frac{G_\mu Q^2}{4\pi\alpha\sqrt{2}}\frac{A_V^p+A_s^p+A_A^p}{\eps\,\GEp^2+\tau\,\GMp^2}\,,
\label{eq:ALRp}\\
A_V^p &=& \xi_V^p \left(\eps\,\GEp^2+\tau\,\GMp^2 \right)              \non\\
      && + \xi_V^n \left( \eps\, \GEp \GEn + \tau\, \GMp \GMn \right)\,, \\
A_s^p &=& \eps\, \GEp\, \xi_V^0 G_E^s + \tau\, \GMp\, \xi_V^0 G_M^s\,, \\
A_A^p &=& - (1-4\shat^2)\sqrt{1-\eps^2}\sqrt{\tau(1+\tau)}\GMp\GAEWp\,.
\end{eqnarray}
The kinematic variables are defined by $\eps \equiv \left[ 1 + 2
(1+\tau) \tan^2{\theta}/{2} \right]^{-1}$ and $\tau \equiv {|Q^2|}/{4
M_p^2}$.  The Standard Model parameters $\alpha$, $G_\mu$ and 
$\shat^2\equiv\sin^2\hat\theta_W$ are taken from the PDG
\cite{Eidelman:2004wy}.
The vector radiative correction factors are defined by $\xi_V^p =
(1-4\shat^2)(1+R_V^p)$, $\xi_V^n = -(1+R_V^n)$ and
$\xi_V^0=-(1+R_V^{(0)})$, with $R_V^p=-0.04471$ and
$R_V^n=R_V^{(0)}=-0.01179$ \cite{Eidelman:2004wy}. The axial radiative
and anapole corrections remain implicit in $\GAEWp$, as this entire
contribution is to be fit to data.

\begin{table*}
\caption{Displayed are the $\eta_i$, appearing in
Eq.~(\ref{eq:APVtheory}), which describe the theoretical asymmetry for
each experiment (in parts-per-million). The measured asymmetry is
shown by $A^{\rm phys}$ and the corresponding uncertainty, $\delta A$,
where sources of error have been added in quadrature. The second
uncertainty, $\delta A_{\rm cor}$, represents the correlated error in
the G0 experiment \cite{Armstrong:2005hs}.
Columns on the right show the determination of the form factors at
$Q^2=0.1\gev^2$ for fits which include all data up to the given
measurement (statistical uncertainty on final decimal place shown in
parentheses). The reduced $\chi^2$ for each fit is displayed, followed in
the final column by the confidence level (CL) for the true value of
the strangeness form factors to be nonzero.
\label{tab:data}}
\begin{ruledtabular}
\begin{tabular}{lr|rrrrr|rrr||rrrrrr}
Collaboration & $Q^2$ & $\eta_0$ & $\eta_A^p$ & $\eta_A^n$ & $\eta_E$ & $\eta_M$ & $A^{\rm phys}$ & $\delta A$ & $\delta A_{\rm cor}$ & $\tilde{G}_A^p$ & $\tilde{G}_A^n$ & $G_E^s$ & $G_M^s$ & $\chi^2$ & CL \\
\hline
SAMPLE   & $0.038$ & $-2.13$ & $0.46$ & $-0.30$ & $1.16$  & $0.28$  & $-3.51$  & $0.81$  & $0$    & ---        & ---        & ---          & ---        & ---   \\
SAMPLE   & $0.091$ & $-7.02$ & $1.04$ & $-0.65$ & $1.63$  & $0.77$  & $-7.77$  & $1.03$  & $0$    & ---        & ---        & ---          & ---        & ---   \\
HAPPEx   & $0.091$ & $-7.50$ & $0$    & $0$     & $-20.2$ & $0$     & $-6.72$  & $0.87$  & $0$    & ---        & ---        & ---          & ---        & ---   \\
HAPPEx   & $0.099$ & $-1.40$ & $0.04$ & $0$     & $9.55$  & $0.76$  & $-1.14$  & $0.25$  & $0$    & ---        & ---        & ---          & ---        & ---   \\
SAMPLE   & $0.1$   & $-5.47$ & $1.58$ & $0$     & $2.11$  & $3.46$  & $-5.61$  & $1.11$  & $0$    & $-2.6(21)$ & $-0.6(30)$ & $-0.044(47)$ & $1.00(75)$ & $1.0$ & 63 \\
PVA4     & $0.108$ & $-1.80$ & $0.26$ & $0$     & $10.1$  & $1.05$  & $-1.36$  & $0.32$  & $0$    & $-2.0(20)$ & $0.3(29)$  & $-0.025(43)$ & $0.87(74)$ & $1.0$ & 71 \\
G0       & $0.122$ & $-1.90$ & $0.06$ & $0$     & $12.0$  & $1.18$  & $-1.51$  & $0.49$  & $0.18$ & $-1.8(19)$ & $0.5(27)$  & $-0.023(43)$ & $0.79(69)$ & $0.7$ & 76 \\
G0       & $0.128$ & $-2.04$ & $0.06$ & $0$     & $12.6$  & $1.30$  & $-0.97$  & $0.46$  & $0.17$ & $-2.4(18)$ & $-0.1(26)$ & $-0.027(42)$ & $0.99(65)$ & $0.7$ & 96 \\
G0       & $0.136$ & $-2.24$ & $0.07$ & $0$     & $13.5$  & $1.48$  & $-1.30$  & $0.45$  & $0.17$ & $-2.5(17)$ & $-0.2(26)$ & $-0.028(42)$ & $1.03(63)$ & $0.6$ & 99 \\
G0       & $0.144$ & $-2.44$ & $0.08$ & $0$     & $14.3$  & $1.67$  & $-2.71$  & $0.47$  & $0.18$ & $-1.6(16)$ & $0.8(25)$  & $-0.021(42)$ & $0.71(61)$ & $1.4$ & 91 \\
G0       & $0.153$ & $-2.68$ & $0.09$ & $0$     & $15.3$  & $1.89$  & $-2.22$  & $0.51$  & $0.21$ & $-1.4(16)$ & $1.0(25)$  & $-0.020(42)$ & $0.66(60)$ & $1.2$ & 91 \\
G0       & $0.164$ & $-2.97$ & $0.11$ & $0$     & $16.5$  & $2.19$  & $-2.88$  & $0.54$  & $0.23$ & $-1.1(16)$ & $1.3(25)$  & $-0.018(42)$ & $0.55(60)$ & $1.2$ & 83 \\
G0       & $0.177$ & $-3.34$ & $0.13$ & $0$     & $18.0$  & $2.58$  & $-3.95$  & $0.50$  & $0.20$ & $-0.4(16)$ & $2.1(24)$  & $-0.012(42)$ & $0.32(59)$ & $1.7$ & 36 \\
G0       & $0.192$ & $-3.78$ & $0.15$ & $0$     & $19.7$  & $3.07$  & $-3.85$  & $0.53$  & $0.19$ & $-0.2(15)$ & $2.3(24)$  & $-0.010(42)$ & $0.24(58)$ & $1.6$ & 18 \\
G0       & $0.210$ & $-4.34$ & $0.19$ & $0$     & $21.8$  & $3.72$  & $-4.68$  & $0.54$  & $0.21$ & $ 0.1(15)$ & $2.7(24)$  & $-0.007(42)$ & $0.14(57)$ & $1.6$ & 1  \\
PVA4     & $0.230$ & $-5.66$ & $0.89$ & $0$     & $22.6$  & $5.07$  & $-5.44$  & $0.60$  & $0$    & $ 0.0(15)$ & $2.5(24)$  & $-0.007(42)$ & $0.14(57)$ & $1.5$ & 1  \\
G0       & $0.232$ & $-5.07$ & $0.23$ & $0$     & $24.4$  & $4.61$  & $-5.27$  & $0.59$  & $0.23$ & $ 0.2(14)$ & $2.8(23)$  & $-0.005(42)$ & $0.09(57)$ & $1.4$ & 3  \\
G0       & $0.262$ & $-6.12$ & $0.31$ & $0$     & $28.0$  & $5.99$  & $-5.26$  & $0.53$  & $0.17$ & $-0.2(14)$ & $2.3(23)$  & $-0.010(41)$ & $0.19(56)$ & $1.4$ & 18 \\
G0       & $0.299$ & $-7.51$ & $0.42$ & $0$     & $32.6$  & $8.00$  & $-7.72$  & $0.80$  & $0.35$ & $ 0.0(14)$ & $2.6(23)$  & $-0.006(41)$ & $0.12(55)$ & $1.3$ & 5  \\
G0       & $0.344$ & $-9.35$ & $0.57$ & $0$     & $38.4$  & $10.9$  & $-8.40$  & $1.09$  & $0.52$ & $ 0.0(14)$ & $2.5(22)$  & $-0.008(41)$ & $0.15(54)$ & $1.2$ & 11 \\
G0       & $0.410$ & $-12.28$& $0.87$ & $0$     & $47.3$  & $16.1$  & $-10.25$ & $1.11$  & $0.55$ & $-0.4(13)$ & $2.1(22)$  & $-0.015(40)$ & $0.27(53)$ & $1.2$ & 44 \\
HAPPEx   & $0.477$ & $-15.46$& $1.12$ & $0$     & $56.9$  & $22.6$  & $-15.05$ & $1.13$  & $0$    & $ 0.1(12)$ & $2.7(21)$  & $-0.004(38)$ & $0.10(49)$ & $1.2$ & 28 
\end{tabular}
\end{ruledtabular}
\end{table*}
Scattering from targets other than the proton provides access to
different flavour components of the nucleon form factors.  The HAPPEx
Collaboration have recently utilised a helium-4 target to directly
extract the strange electric form factor \cite{Aniol:2005zf}, where
the theoretical asymmetry can be written as
\begin{equation}
A_{LR}^{\mathrm{He}} = \frac{G_\mu Q^2}{4\pi\alpha\sqrt{2}}
             \left[ (\xi_V^p+\xi_V^n) + 2\frac{\xi_V^0 G_E^s}{G_E^p+G_E^n}\right]\,.
\end{equation}

In the SAMPLE experiment, which detected electrons scattered at
backward angles, the contribution from $G_E^s$ is substantially
suppressed. These measurements were primarily sensitive to a
linear combination of the axial and strange magnetic form factors. In
addition to the proton target, the PV-asymmetry has also been measured
on the deuteron \cite{Ito:2003mr}. While providing a
different combination of $\GMs$ and $\GAEWp$, this also introduces
sensitivity to the neutron axial form factor.

Scattering from the deuteron is dominated by the quasielastic
interaction with the nucleon constituents.  The analysis of the
deuteron results \cite{Beise:2004py} has also included nuclear
corrections, involving a realistic deuteron wave-function,
rescattering effects and the small contribution from elastic deuteron
scattering~\cite{Schiavilla}. Further parity-violating contributions
arising from the deuteron wavefunction and exchange currents, while
small \cite{Schiavilla}, have been included.

A combined analysis of the current world PV data requires a consistent
treatment of the vector and axial form factors and radiative
corrections. Our theoretical asymmetries have therefore been
reconstructed for each measurement.  The theoretical asymmetry is
\begin{equation}
A_{PV}^{\rm theory} = \eta_0 + \eta_A^p \GAEWp + \eta_A^n \GAEWn + \eta_E \GEs + \eta_M \GMs\, ,
\label{eq:APVtheory}
\end{equation}
where the values of $\eta_i$, given in Table~\ref{tab:data}, include
the latest vector form factors \cite{Kelly:2004hm} and PDG radiative
corrections.

It has been observed that the strange form factors are mildly
sensitive to the choice of form factor parameterisation, with an
uncertainty dominated by the neutron charge form factor. To test the
sensitivity to $\GEn$, we explicitly included the experimental data
for $\GEn$~\cite{Hyde-Wright:2004gh} in our global fit. Over the
low-$Q^2$ domain required in this analysis, the form factor can be
parameterised by a Taylor expansion up to ${\cal O}(Q^6)$. This made
no significant difference to the final extraction, and hence the
central value of the Kelly parameterisation \cite{Kelly:2004hm} is
taken in the following analysis.

%
%
\begin{figure}
\includegraphics[width=\columnwidth]{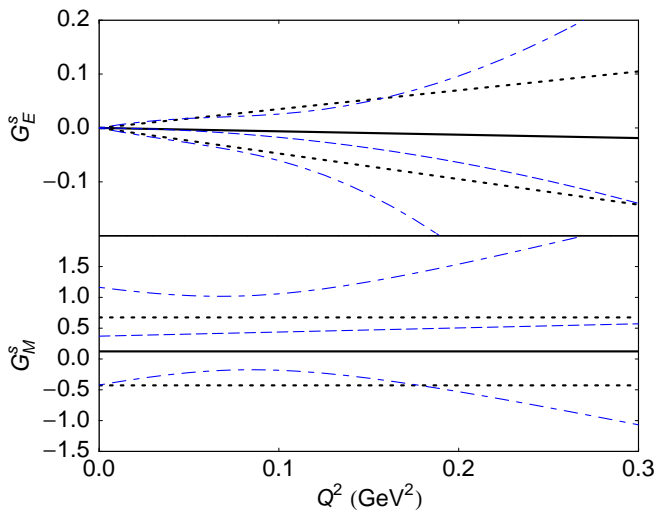}
\caption{The strange electric and magnetic form factors. The solid
curve shows the leading-order fit, with 1--$\sigma$ bound shown by the
dotted curves. The dashed and dash-dotted curves show the 
fit and error of the next-to-leading order fit. 
\label{fig:GVs} }
\end{figure}
In order to extract all three form factors using as much data as
possible, we parameterise their $Q^2$ dependence.  At low momentum
transfer, a Taylor series expansion in $Q^2$ is sufficient and
minimises the model dependence of the determined form factors.  The
quality of a Taylor series expansion can be estimated
phenomenologically. Vector meson dominance would suggest that the
$Q^2$ evolution of the form factors be no more rapid than a dipole
with mass parameter $\sim m_\phi \sim 1\gev$.  Similarly, lattice QCD
simulations in the vicinity of the strange quark yield behaviour
consistent with a dipole of scale $>1\gev$
\cite{Gockeler:2003ay,Ashley:2003sn}.  With the aim of fitting data up
to $Q^2\sim 0.3\gev^2$, approximating a dipole by a constant over this
range would lead to less than 20\% uncertainty (less than 10\% at the
next order in $Q^2$).

To isolate the individual form factors at higher-$Q^2$, a
combination of neutrino and parity-violating electron scattering
should provide the tightest constraint, as described in
Ref.~\cite{Pate:2003rk}.

We describe the $Q^2$-dependence of the form factors over the range
$0<Q^2<0.3\gev^2$ by
\begin{eqnarray}
&\GAEWN  = \tilde{g}_A^N \left( 1 + Q^2/M_A^2 \right)^{-2}\, ,& \label{eq:gAdip} \\
&\GEs = \rho_s Q^2 + \rho_s' Q^4\, , \qquad
\GMs = \mu_s + \mu_s'  Q^2\, .& \label{eq:Taylor}
\end{eqnarray}
The momentum dependence of the radiative corrections is assumed to be
mild, and therefore the axial dipole mass is chosen to be that
determined from neutrino scattering, $M_A=1.026\gev$
\cite{Bernard:2001rs}.

\begin{figure}[!t]
\includegraphics[width=\columnwidth]{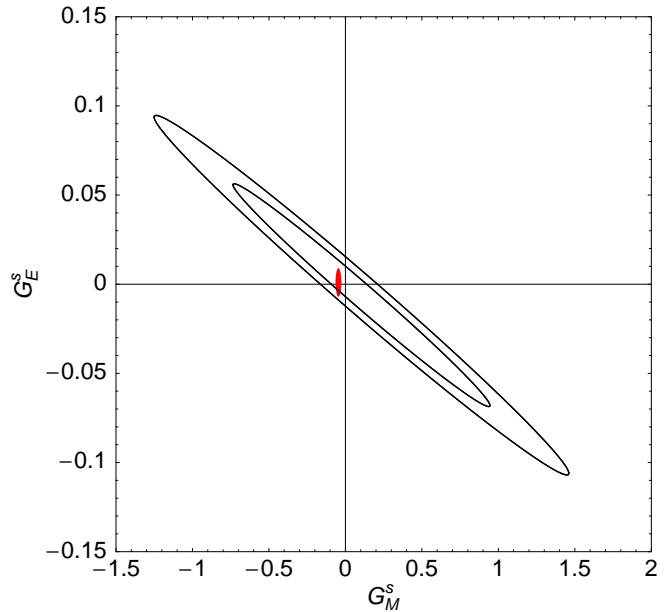}
\caption{
The contours display the the 68 and 95\% confidence intervals for
the joint determination of $\GMs$ and $\GEs$ at $Q^2=0.1\gev^2$.  
The solid ellipse shows the theoretical results of Leinweber
et~al.~\cite{Leinweber:2004tc,Leinweber:2006ug}.
\label{fig:GMsGEs}}
\end{figure}

\begin{figure}
\includegraphics[width=\columnwidth]{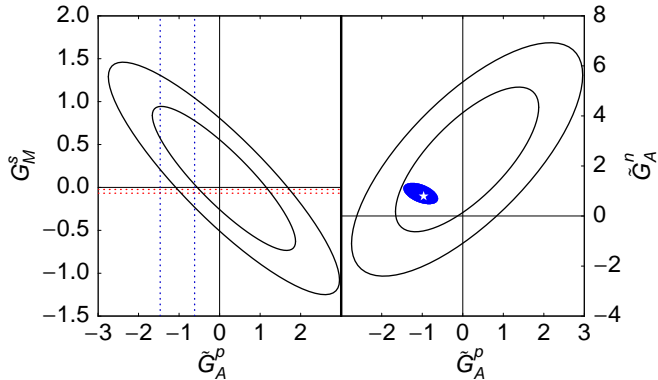}
\caption{
The contours display the the 68 and 95\% confidence intervals
for the joint determination of the form factors (defined on the axes)
at $Q^2=0.1\gev^2$. 
The horizontal and vertical bands in the left panel shows the theory
results of Leinweber~et~al.~\cite{Leinweber:2004tc} and
Zhu~et~al.~\cite{Zhu:2000gn}, respectively.
The disk in the right panel displays the theoretical result of
Zhu~et~al.~\cite{Zhu:2000gn}, with the white star indicating the zero
anapole origin. (The same dipole behaviour as Eq.~(\ref{eq:gAdip}) is
assumed.)
\label{fig:cont2}}
\end{figure}

The best fit for $Q^2<0.3\gev^2$ yields, at leading order in $Q^2$, a
reduced $\chi^2=19.7/15=1.3$, with parameters
\begin{eqnarray}
\tilde{g}_A^p  &=&  0.05 \pm 1.38 \mp 0.29 \, , \\
\tilde{g}_A^n  &=&  2.61 \pm 2.27 \mp 0.37 \, , \\
\rho_s         &=& -0.06 \pm 0.41 \mp 0.00 \gev^{-2} \, , \\
\mu_s          &=&  0.12 \pm 0.55 \pm 0.07 \, .
\end{eqnarray}
The second error bar displays the sensitivity to the
correlated error
in the G0 experiment, where the data
has been refit using $A^{\rm phys}\pm\delta A_{\rm cor}$. The 
extraction of the strange form factors over the low-$Q^2$ range is
shown in Fig.~\ref{fig:GVs}. We display the joint determination of the
strange electric and magnetic form factors at $Q^2=0.1\gev^2$ in
Fig.~\ref{fig:GMsGEs}, where we also show the theoretical calculations
of Leinweber et al.~\cite{Leinweber:2004tc,Leinweber:2006ug}. Similar
contours in $\tilde{G}_A^p$--$G_M^s$ and
$\tilde{G}_A^p$--$\tilde{G}_A^n$ space are shown in
Fig.~\ref{fig:cont2}.

The stability of the fits to truncation of the data set at a maximum
$Q^2$ value has been investigated. The resulting fits are displayed in
Table~\ref{tab:data}, where a clear signal for nonzero strangeness is
observed in the vicinity of $Q^2\sim 0.1\gev^2$ --- with caution that
the fits are particularly sensitive to truncation up until
$Q^2\sim0.2\gev^2$. To investigate a potential enhancement near
$Q^2\sim 0.1\gev^2$, we include the second-order terms of
Eq.~(\ref{eq:Taylor}) and fit all data for $Q^2<0.3\gev^2$.
This produces a $\chi^2=18.1/13=1.4$ and best-fit parameters
$\tilde{g}_A^p=-0.80 \pm 1.68$, $\tilde{g}_A^n=1.65\pm2.62$,
$\rho_s=-0.03\pm0.63\gev^{-2}$, $\rho_s'=-1.5\pm5.8\gev^{-4}$,
$\mu_s=0.37\pm0.79$ and $\mu_s'=0.7\pm6.8\gev^{-2}$, where the errors are
statistical only. 
Figure~\ref{fig:GVs} shows the uncorrelated separation of the electric
and magnetic form factors at this order. Where the data is best
constrained, $Q^2\sim 0.1\gev^2$, there is only a 55\% CL in support
of nonzero strangeness. This suggests that the strangeness signal in
Table~\ref{tab:data}, obtained by truncating the data at $Q^2\sim
0.14\gev^2$, is consistent with a random fluctuation.

%
%
Previous (non-global) attempts to extract the nucleon strange form
factors from world data used a theoretical prediction of
$\tilde{g}_A^N$ \cite{Zhu:2000gn}. In the following, we compare the
axial form factors extracted from the data with this prediction. We
write the axial charges, Eq.~(\ref{eq:gAdip}), as
\begin{equation}
\tilde{g}_A^N = \xi_A^{T=1} \GA \tau_3 + \xi_A^{T=0} a_8 + \xi_A^{0} a_s + A_{\rm ana}^N\, ,
\label{eq:gA}
\end{equation}
with $\tau_3=1(-1)$ for the proton (neutron). The radiative
corrections are implied to be single-quark only $\xi_A^{T=1}=-0.828$,
$\xi_A^{T=0}=-0.126$ and $\xi_A^{0}=0.449$ \cite{Eidelman:2004wy}. 
The axial charges are relatively well known, where we use
$\GA=1.2695$, $a_8=0.58\pm
0.03\pm 0.12$ \cite{Filippone:2001ux} and $a_s=-0.07\pm 0.04\mp 0.05$
\cite{Adams:1997tq}. The second error in $a_8$ and $a_s$ reflects
estimates of the SU(3)-flavour symmetry violations of $\sim$20\% in
the determination of $a_8$ from hyperon $\beta$-decay
\cite{Zhu:2000zf}. 
The dominant source of uncertainty in Eq.~(\ref{eq:gA}) is the anapole
contribution, $A_{\rm ana}^N=A_{\rm ana}^{(T=1)}\tau_3+A_{\rm ana}^{(T=0)}$.
Converting the result of Zhu~et~al.~\cite{Zhu:2000gn} to
$\overline{MS}$ \cite{Kumar:2000eq}, the anapole terms are estimated
to be $A_{\rm ana}^{(T=1)}=-0.11\pm0.44$ and
$A_{\rm ana}^{(T=0)}=0.02\pm0.26$.
This gives the total theory estimates for the axial
charges in PVES, 
$\tilde{G}_A^p = (-1.16\pm0.04)+(-0.09 \pm 0.51)$
and $\tilde{G}_A^n = ( 0.95\pm0.04)+( 0.13 \pm 0.51)$,
where the second term is the anapole contribution. These estimates are
consistent with the present determination, as shown in the right panel
of Fig.~\ref{fig:cont2}.

As we see from Table~\ref{tab:data}, the current world data is
consistent with the strange form factors being zero at a high level of
confidence. 
The anapole contributions,
considered alone, are also consistent with zero.  On the other hand,
if one interrogates the data for the probability that strange and
anapole form factors are simultaneously zero, within the current
errors, the hypothesis is only supported at 8\%. While the present
data set cannot distinguish the origin of this effect, there appears
to be significant support for a nonzero signal in at least one of the
strange or anapole contributions.

%
%
In conclusion, our analysis of the world data set for PVES has yielded
the best experimental determination, at low $Q^2$, of the strange
electric and magnetic form factors of the proton as well as the anapole
form factors of the proton and neutron.
While both the strangeness and anapole contributions are consistent
with zero, we expect that the additional HAPPEx and G0-backward angle
experiments at Jefferson Lab and the PVA4-backward angle experiment at
Mainz will soon yield data that, when combined with this analysis,
could reveal a nontrivial result for at least one of these form
factors.

%
%
We wish to express our gratitude to M.~Pitt, D.~Bowman, K.~de~Jager,
W.~Melnitchouk, M.~Paris, K.~Paschke and R.~Schiavilla for helpful
discussions. This work was supported by DOE contract
DE-AC05-84ER40150, under which SURA operates Jefferson Lab.

%
%

\end{document}